\documentclass[conference]{IEEEtran}
\IEEEoverridecommandlockouts
\usepackage{cite}
\usepackage{amsmath,amssymb,amsfonts}
\usepackage{soul}
\usepackage{algorithmic}
\usepackage{graphicx}
\usepackage{textcomp}
\usepackage{xcolor}
\usepackage{amsfonts}
\usepackage{booktabs}
\usepackage{float}
\usepackage{graphicx}
\usepackage{grffile}
\usepackage{longtable}
\usepackage{wrapfig}
\usepackage{gensymb}
\usepackage{textcomp}
\usepackage{multicol}
\usepackage{amsmath}
\usepackage{algorithm}
\usepackage{verbatim}
\usepackage{graphicx}
\usepackage{caption}
\usepackage{listings}
\usepackage{array}
\usepackage{subcaption}
\usepackage{booktabs}
\usepackage{amsmath}
\usepackage{enumitem}
\usepackage{dblfloatfix}
\usepackage{balance}
\usepackage{lastpage}
\usepackage{lipsum}
\usepackage[utf8]{inputenc}

\begin{document}
\title{Blockchain-Enhanced Framework for Secure Third-Party Vendor Risk Management and Vigilant Security Controls}
\author{\IEEEauthorblockN{Deepti Gupta\IEEEauthorrefmark{1}, Lavanya Elluri\IEEEauthorrefmark{2}, Avi Jain\IEEEauthorrefmark{3}, Shafika Showkat Moni\IEEEauthorrefmark{4}, Omer Aslan\IEEEauthorrefmark{5}}
\IEEEauthorblockA{\IEEEauthorrefmark{1}\IEEEauthorrefmark{2}Dept. of Computer Information Systems, Texas A\&M University - Central Texas, Texas, USA \\\IEEEauthorrefmark{3}{College of Business, University of Tampa, Florida, USA}\\\IEEEauthorrefmark{4}Dept. of Electrical Engineering and Computer Science, Embry-Riddle Aeronautical University, Florida, USA\\\IEEEauthorrefmark{5} Department of Software Engineering, Bandırma Onyedi Eylül University, Bandırma, Turkey\\}
\IEEEauthorrefmark{1}d.gupta@tamuct.edu, 
\IEEEauthorrefmark{2}elluri@tamuct.edu,
\IEEEauthorrefmark{3}avi.jain@spartans.ut.edu,
\IEEEauthorrefmark{4}monis@erau.edu,
\IEEEauthorrefmark{5}oaslan@bandirma.edu.tr}

\maketitle

\begin{abstract}

In an era of heightened digital interconnectedness, businesses increasingly rely on third-party vendors to enhance their operational capabilities. However, this growing dependency introduces significant security risks, making it crucial to develop a robust framework to mitigate potential vulnerabilities. This paper proposes a comprehensive secure framework for managing third-party vendor risk, integrating blockchain technology to ensure transparency, traceability, and immutability in vendor assessments and interactions. By leveraging blockchain, the framework enhances the integrity of vendor security audits, ensuring that vendor assessments remain up-to-date and tamper-proof. This proposed framework leverages smart contracts to reduce human error while ensuring real-time monitoring of compliance and security controls. By evaluating critical security controls—such as data encryption, access control mechanisms, multi-factor authentication, and zero-trust architecture—this approach strengthens an organization's defense against emerging cyber threats. Additionally, continuous monitoring enabled by blockchain ensures the immutability and transparency of vendor compliance processes. In this paper, a case study on iHealth's transition to AWS Cloud demonstrates the practical implementation of the framework, showing a significant reduction in vulnerabilities and marked improvement in incident response times. Through the adoption of this blockchain-enabled approach, organizations can mitigate vendor risks, streamline compliance, and enhance their overall security posture. Our findings highlight the importance of employing blockchain to enforce security controls and maintain compliance with healthcare regulations such as HIPAA. In this paper, we present a comprehensive set of security controls and demonstrate how blockchain technology enhances their effectiveness, ensuring greater transparency, accountability, and automation in vendor assessments. By reducing human error, enabling real-time monitoring, and validating compliance, blockchain strengthens the overall security and resilience of the third-party vendor ecosystem.

\end{abstract}

\begin{IEEEkeywords}
Vendor Assessment, Third Party, Security and Privacy, Threats, Risks and Attacks, Blockchain.
\end{IEEEkeywords}

\section{Introduction}
\begin{figure}
\centering
\includegraphics[width=.5\textwidth, height=.3
\textheight]{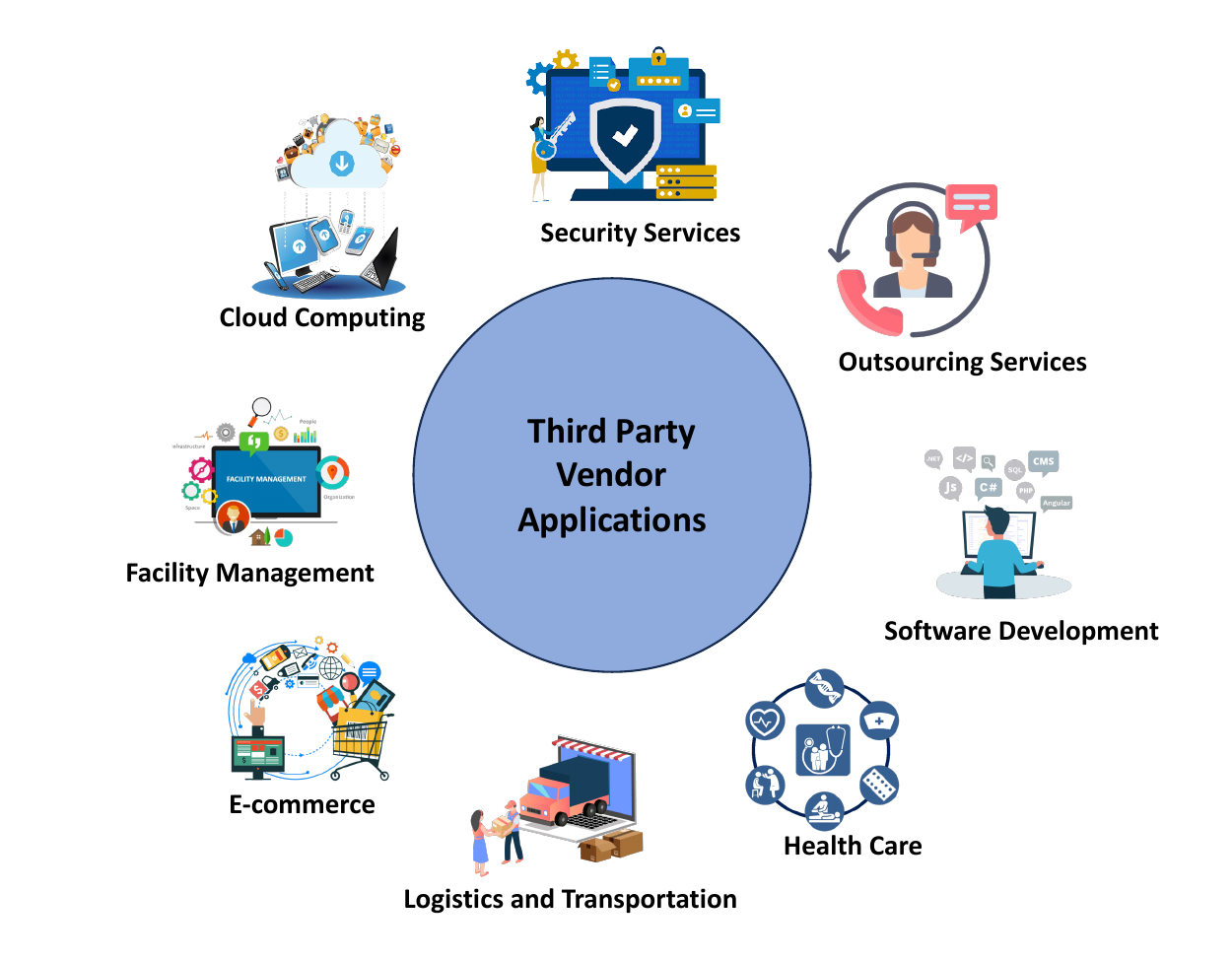}
\centering
\caption{Third-Party Vendor Applications}
\label{application}
\end{figure}

In a time of increased digital connectivity, businesses often turn to third-party vendors to boost their operational capabilities. Although the usage of third-party vendors in all sectors, namely e-commerce, health care, IT services, cloud computing, security services, and software development (Fig. \ref{application}), is growing, this reliance introduces considerable security risks, highlighting the need for a solid framework to address and minimize potential vulnerabilities. In recent years, notable third-party vendor security breaches have affected companies like Uber, Target, Equifax, Marriott, and SolarWinds. Sensitive data, including credit card numbers, social security numbers, and personal identification information, has been compromised by security attacks, putting the personal data of millions at risk. Uber announced a third-party data breach in December 2022\footnote{https://cybernews.com/news/uber-suffers-data-breach-attack-third-party-vendor/}. The hackers published employee email addresses, IT asset details, and corporate report data online for Uber and Uber Eats. The hackers accessed this information through Teqtivity, a vendor providing the company's tech and IT asset-tracking solutions. Security experts warn that the leaked data has sufficient details to launch targeted phishing attacks on Uber employees. These threats underscore the risks linked to third-party vendors, emphasizing the importance of robust cybersecurity controls due to the failure to follow proper assessment protocols. 

This study~\cite{kiesel2022analyzing} shows that a similar multi-vector ransomware attack happened on the Accellion File Transfer Appliance server. In this attack, the appliance's security was compromised, allowing malicious actors to gain unauthorized access and copy data. The attackers threatened to release the data unless a ransom was paid. To prevent this, Accellion and affected client companies collaborated to provide defense solutions, customer support, and guidance on handling the ransom demand. The attack significantly impacted Accellion's clients, such as the Federal Reserve Bank of New Zealand, which incurred expenses of 3.5 million dollars to address the incident. Additionally, the bank had to allocate 17,500 hours of internal resources to assist in responding to the attack. It sheds light on the critical role of collaboration between organizations and their solution providers when facing such threats. Furthermore, it underscores the importance of proactively implementing security measures such as routine updates, rigorous vetting of third-party vendors, and automatic updates to mitigate the risk of future attacks.

In current scenario, large multi-national corporations decide to migrate its data infrastructure to the cloud, specifically leveraging services provided by third party services such as Amazon Web Services (AWS), Google Cloud Platform (GCP) and Snowflake for data storage and analytics. Vendor assessment are crucial for security, data protection, reliability, performance and regulatory compliance etc. In terms of security and privacy, it is required to conduct a thorough vendor assessment that helps to ensure that these vendors adhere to industry best practices, compliance standards (such as GDPR or HIPAA), and have robust security controls in place to safeguard against cyber threats and unauthorized access. Overall, conducting a comprehensive vendor assessment of third-party providers like AWS, GCP and Snowflake is essential for mitigating risks, ensuring compliance, optimizing costs, and maximizing the value proposition of cloud-based data infrastructure solutions for large corporations.

The majority of organizations adhere to National Institute of Standards and Technology (NIST) guidelines, which provide a standardized set of security controls for assessing third-party vendors. Integrating NIST frameworks, such as the NIST Cybersecurity Framework~\cite{NIST_cy}, SP 800-53 Rev. 5~\cite{NIST_SP}, and SP 800-161 Rev. 1~\cite{R}, allows organizations to systematically evaluate and mitigate supplier cybersecurity risks. These guidelines focus on key aspects like contractual restrictions, regular assessments, and supply chain risk management. As third-party risk management becomes increasingly crucial for addressing the complexities of modern supply chains, adopting these practices ensures a proactive approach to identifying and mitigating cybersecurity threats. However, relying solely on NIST may not provide a comprehensive and robust security framework for vendor assessment, potentially leading to overlooked vulnerabilities and errors.

To further enhance the security of third-party vendor risk management, proposed framework incorporates blockchain technology as a critical component. Blockchain’s decentralized ledger system provides a transparent and tamper-proof way to track transactions, contracts, and interactions with third-party vendors, improving traceability and accountability. By using blockchain, organizations can ensure the integrity of vendor assessments and establish immutable audit trails, reducing the risk of fraud, tampering, or unauthorized access. This blockchain-based approach strengthens trust between organizations and their third-party vendors, enabling more secure and resilient partnerships. Developing a comprehensive framework that integrates blockchain technology alongside security controls is imperative for organizations seeking to mitigate the vulnerabilities associated with third-party vendor relationships.

This study aims to offer actionable insights that help organizations navigate the complex landscape of cybersecurity challenges and foster a culture of security within their software development processes by addressing the intersection of NIST compliance guidelines and third-party risk management. The ultimate goal is to empower vendor organizations to establish a secure software development lifecycle, thereby enhancing their overall cybersecurity posture and resilience against the relentless evolution of cyber threats. The main contribution of this paper are as follows-
\begin{itemize}

\item In our research, we have diligently identified a critical gap pertaining to third-party vendor risks. Through thorough analysis and assessment, we've uncovered vulnerabilities and potential threats.

\item We enhance the NIST framework for third-party vendor risks by meticulously integrating a comprehensive array of security controls tailored to address the specific vulnerabilities inherent in third-party relationships.

\item We introduce a secure framework tailored for third-party vendors and implement this innovative framework with security controls, including a blockchain-based approach, to ensure transparency and immutability. 

\item We conduct a comprehensive case study on a smart medical device vendor to meticulously identify and analyze potential security vulnerabilities and risks.
 \end{itemize}

The remainder of this paper is organized as follows. Section~\ref{related} presents the literature review on third-party vendor security risk and control. We discuss the NIST security framework in Section~\ref{nist}. Section~\ref{security controls} presents the security threats, challenges associated with those threats, countermeasures, and security controls that could be implemented to stay secure from those emerging threats, illuminating the path toward a secure Systems development life cycle (SDLC). The blockchain-enhanced framework is discussed in the Section~\ref{framework}. Section~\ref{study} presents the case study on a smart medical device vendor to identify security issues. Section~\ref{results} presents the results. Conclusion and future work are discussed in Section~\ref{conclusion}.

\section{Related Work and Background}
\label{related} 

In recent years, attacks targeting organizations have evolved into significant threats, posing substantial risks to the security and integrity of systems and networks. These attacks, often orchestrated by sophisticated threat actors, have become increasingly sophisticated and diverse. Specifically, attacks against third-party vendors have emerged as a major concern, employing various techniques to compromise security defenses and gain unauthorized access to sensitive information or systems. Attackers exploit vulnerabilities in vendor software, manipulate supply chain processes, or compromise vendor credentials to infiltrate organizations. The consequences of such breaches can be severe, leading to data loss, financial damage, reputational harm, and regulatory penalties. Multi-vector ransomware attacks present a formidable challenge for cybersecurity professionals. These attacks combine multiple techniques to breach systems and threaten sensitive data unless a ransom is paid. Such attacks complicate the defense landscape, requiring organizations to guard against multiple entry points simultaneously to prevent data exfiltration and extortion.

Research highlights the importance of regulatory compliance in selecting secure services. For example, Mollakuqe et al.\cite{mollakuqe2023privacy} emphasized the role of GDPR in service selection, proposing a security classification system for Low, Medium, and High-security levels to guide decision-making. Khan et al.\cite{khan2021analyzing} conducted a systematic literature review identifying critical security challenges for software vendors using cloud-based big data platforms. They found recurring challenges like data secrecy, unauthorized access, and trust issues, recommending practices to enhance vendor security. Similarly, Wu et al.~\cite{wu2019cloud} proposed a game-theory approach to evaluate the trustworthiness of public cloud providers and mediators, helping users assess privacy risks and improve security strategies. Despite these efforts, a gap remains in providing a comprehensive framework to assess and manage third-party vendor security risks across domains such as cloud computing, big data, smart healthcare, and smart cities. Existing studies largely focus on isolated aspects like data confidentiality and access control, but a holistic approach is needed to tackle the diverse challenges. In addition, several security models for protecting IoT devices are discussed in~\cite{gupta2020access, gupta2021hierarchical, gupta2023integration, gupta2022game, kotal2023privacy, kayode2020towards}.

Blockchain technology offers a promising solution to the challenges of security and privacy. Yaqoob et al.\cite{yaqoob2022blockchain} and Kummar et al.\cite{kummar2022blockchain} demonstrated that blockchain can enhance trust and security in third-party transactions, especially in supply chain and healthcare environments where data integrity and auditability are critical. However, their work also revealed significant limitations, such as inadequate frameworks for comprehensive third-party risk assessment, reliance on potentially outdated security standards, and challenges in integrating blockchain with existing compliance protocols. Additionally, scalability issues and inefficiencies in real-time processing further hinder the effectiveness of these solutions in managing third-party risks effectively.

Based on the literature survey, the primary gap in vendor security risk management lies in the absence of a comprehensive and standardized approach to assess and mitigate the security risks posed by third-party vendors across various domains, including cloud computing, big data, smart healthcare, and smart cities. While previous studies have addressed specific aspects such as data confidentiality, access control, and trustworthiness evaluation, a more integrated framework is necessary to tackle the diverse security challenges across these vendor environments. An integrated framework must incorporate guidelines, standards, and best practices to help organizations effectively evaluate, mitigate, and manage third-party vendor risks. This framework should ensure the protection of sensitive data, system integrity, and secure collaboration between vendors and organizations at all operational levels. To bridge this gap, blockchain technology can play a transformative role. By leveraging blockchain's decentralized and immutable ledger, organizations can achieve enhanced transparency, traceability, and security in third-party vendor relationships. Blockchain ensures that all transactions and interactions between organizations and vendors are securely recorded and verified, reducing the risk of tampering, unauthorized access, and data breaches. Additionally, smart contracts can automate compliance checks and enforce security policies, ensuring vendors adhere to agreed-upon security standards and regulatory requirements. Implementing blockchain-based solutions for third-party vendor management will significantly strengthen trust, accountability, and data integrity, offering a robust mechanism to handle the evolving complexities of vendor risks.

\section{NIST Framework}
\label{nist} 

NIST~\cite{prevalent} sets IT-related standards and guidelines widely adopted by both federal agencies and over 50\% of private sector organizations. Key NIST publications relevant to third-party risk management include the Cybersecurity Framework (CSF) v1.1, SP 800-53 Rev. 5, and SP 800-161 Rev. 1. Together, these guidelines offer a comprehensive approach to assessing and mitigating supplier cybersecurity risks. This guideline covers critical aspects such as contractual restrictions, regular inspections, risk management plans, and response/recovery strategies, making them essential tools for managing vendor relationships. SP 800-53~\cite{Force_2020} emphasizes supplier evaluation and third-party solutions, with recent revisions reflecting an evolution in NIST’s Supply Chain Risk Management (SCRM) methodology. The newly added ``SR-Supply Chain Risk Management" control group in Rev. 5 focuses on formal risk plans, collaboration, and transparency. SP 800-161 Rev. 1 complements this by providing a detailed framework for managing supply chain cybersecurity risks. Together with SP 800-53, these guidelines establish a strong foundation for assessing and mitigating supplier risks effectively. Additionally, NIST’s Cybersecurity Framework v1.1~\cite{NIST_2023} integrates established frameworks (CIS, COBIT, ISA, ISO/IEC) and outlines procedures for identification, risk assessments, prioritizing suppliers, contractual measures, and regular reviews, making it highly effective for third-party risk and supply chain security.

NIST framework emphasizes several key aspects of vendor risk management. Organizations must start with comprehensive risk assessments to identify and evaluate cybersecurity risks posed by third-party vendors. Rigorous vendor selection criteria should be established to ensure potential vendors meet security standards and compliance requirements, reducing risks from the outset. Clear contractual agreements are vital, detailing each party's security responsibilities, including data protection, incident response, and regulatory compliance. Ongoing monitoring of vendor activities ensures the detection of any security incidents or contractual deviations, allowing for timely intervention and remediation. However, organizations still need to tailor a robust framework to address specific challenges associated with third-party vendors, such as setting clear contractual requirements, conducting thorough assessments, and implementing continuous monitoring to ensure compliance.

While NIST’s framework is invaluable, implementing it poses challenges. The complexity of cybersecurity, the diverse range of organizational needs, an evolving threat landscape, and reliance on vendor compliance all create difficulties. To overcome these, organizations must adopt adaptive and proactive approaches to vendor risk management. In the next section, we discuss the essential security controls necessary for third-party assessments and how they can be incorporated into the NIST framework.
\section{Essential Security Controls for Third-Party Assessment}
\label{security controls} 

\begin{figure*}[t]
\centering
 \includegraphics[width=1\textwidth, height=.40\textheight]{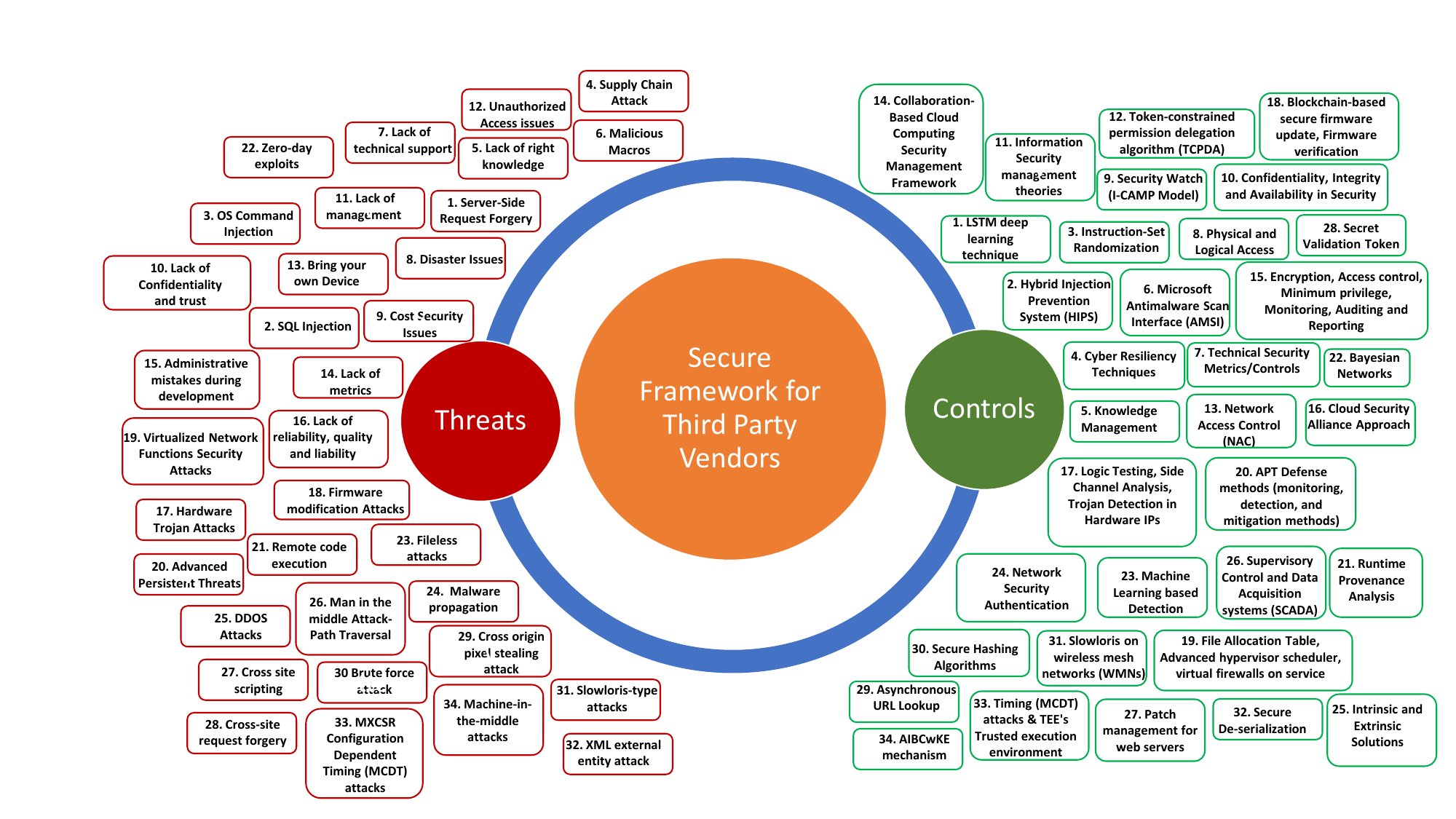}
\centering
\caption{Essential Security Controls for Third-Party Assessment}
\label{vendor}
\end{figure*}

In this section, we examine significant security threats that third-party vendors face, the associated challenges, and the implementation of appropriate countermeasures to enhance vendor resilience. We present a secure framework (Fig. \ref{vendor}) that addresses these threats, offering practical control measures to bolster security. This framework can be applied universally by organizations to manage risks effectively. 

\begin{itemize}

\item Server-Side Request Forgery (SSRF):
A critical threat, SSRF allows attackers to manipulate server behavior to access internal systems. To counter this, Long Short-Term Memory (LSTM) models are employed as a deep learning technique~\cite{al2021detecting}. LSTM excels at detecting patterns in server requests, which enhances the system's ability to identify and neutralize suspicious requests, strengthening the server’s security posture.

\item  SQL Injection:
SQL Injection remains a prevalent threat to databases. The Hybrid Injection Prevention System (HIPS) offers a robust solution by combining signature-based and anomaly-based detection to thwart malicious queries~\cite{makiou2014improving}. This dual approach helps safeguard sensitive data by preventing unauthorized database manipulation.

\item OS Command Injection:
To mitigate OS Command Injection attacks, Instruction-Set Randomization dynamically alters the instruction set architecture, making it significantly harder for attackers to exploit predefined commands~\cite{kc2003countering}. This enhances system resilience by introducing unpredictability in command execution.

\item Supply Chain Attacks:
Supply chain vulnerabilities expose organizations to third-party risks. Implementing Cyber Resilience Techniques like code reviews and secure development practices can fortify the supply chain against threats~\cite{heinbockel2017supply}, ensuring integrity and reducing the risk of breaches.

\item Malicious Macros:
A common vector for attacks is the execution of malicious macros in widely used software. Microsoft Antimalware Scan Interface (AMSI) provides a real-time defense by scanning and evaluating macros during runtime, detecting malicious behavior before execution~\cite{chen2023malicious}.

\item Unauthorized Access:
Preventing unauthorized access is critical for maintaining system integrity. The Token-Constrained Permission Delegation Algorithm (TCPDA) helps limit user privileges by controlling access through tokens~\cite{shi2020mechanism}. This restricts access rights based on task-specific needs, mitigating risks from excessive permissions.

\item Bring Your Own Device (BYOD):
The growing trend of BYOD introduces security risks through personal devices. Network Access Control (NAC) ensures that only secure, authorized devices can connect to the network, protecting against unauthorized access and potential malware spread~\cite{rivera2013analysis}.

\item Firmware Modification Attacks:
Firmware manipulation can lead to severe vulnerabilities in embedded systems. Blockchain-based Secure Firmware Update and Firmware Verification mechanisms ensure firmware integrity, preventing unauthorized modifications~\cite{bettayeb2019firmware}. These methods offer transparency and traceability, critical in securing firmware updates.

\item Malware Propagation:
Malware can spread rapidly across networks, jeopardizing security. Network Security Authentication, including multi-factor authentication, adds a vital layer of defense by ensuring that only legitimate users and devices can access network resources~\cite{antrosiom2005malware}.

\item Distributed Denial of Service (DDoS) Attacks:
DDoS attacks disrupt services by overwhelming networks with traffic. A combination of intrinsic and extrinsic solutions is effective. Intrinsic methods like traffic filtering protect the internal network, while extrinsic services, such as cloud-based DDoS protection, help absorb attack traffic~\cite{eliyan2021and}.

\item Zero-Day Exploits:
Bayesian Networks offer a proactive defense against Zero-Day attacks. By analyzing system behaviors and prioritizing threats through probabilistic modeling, these networks detect novel exploits, enhancing overall system resilience~\cite{li2017effective}.

\item Fileless Attacks:
Fileless attacks evade traditional malware defenses. Machine Learning-based Detection offers an adaptive approach, identifying abnormal behavior patterns indicative of such attacks~\cite{liu2023survey}. This allows organizations to respond swiftly to emerging threats.

\item Advanced Persistent Threats (APTs):
APTs are sophisticated, long-term attacks targeting critical assets. APT Defense Methods, including real-time monitoring and advanced threat detection techniques, provide continuous visibility and rapid response to neutralize APTs before they cause significant harm~\cite{alshamrani2019survey}.

\item Unauthorized Remote Code Execution (RCE):
RCE allows attackers to execute malicious code on a remote system. Runtime Provenance Analysis tracks the origin of processes in real-time, mitigating unauthorized code execution~\cite{xiao2022understanding}.

\item Command Injection (Path Traversal):
In the context of Supervisory Control and Data Acquisition (SCADA) systems, command injection attacks pose a significant threat. Implementing strong input validation and regular security assessments helps mitigate the risks of such attacks~\cite{upadhyay2020scada}.

\end{itemize}

This exploration highlights how prevalent security issues among third-party vendors expose organizations to significant risks, emphasizing the importance of robust security controls to address these vulnerabilities. Threats continually adapt, often targeting supply chain partners with inadequate security measures. These threats exploit weaknesses in vendor networks, which can lead to unauthorized access, data leaks, and data breaches. To mitigate such risks, organizations must identify vulnerabilities common to third-party environments, including weak access controls, data exposure risks, and insufficient encryption. Tackling these issues demands a multifaceted approach: regular security assessments help detect gaps before they become threats, while continuous monitoring ensures any abnormal activity is promptly flagged. Additionally, blockchain-based auditing provides an immutable and transparent record of vendor transactions and access logs, enabling traceable accountability across the supply chain. Together, these measures not only mitigate risk but also foster greater transparency and trust among all parties involved.

\section{Blockchain-Enhanced Framework}
\label{framework}

The proposed blockchain-enhanced framework for third-party vendor assessment leverages the inherent properties of blockchain technology to establish a transparent, secure, and immutable system for managing vendor relationships and assessing associated risks. By integrating blockchain into the vendor assessment process, organizations can create a decentralized ledger that records all vendor interactions, compliance checks, and risk assessments. This transparency allows stakeholders to verify the authenticity of vendor credentials and certifications, ensuring that only compliant vendors are engaged. There are several benefits to use this technology for vendor assessment including enhanced trust and transparency, improved risk management, streamlined processes, cost efficiency, stronger vendor relationships, and adaptability to evolving threats. The use of blockchain technology provides a high level of transparency, allowing all stakeholders to access a consistent and verifiable record of vendor interactions and compliance status. This transparency fosters trust between organizations and their vendors, as each party can independently verify the other’s claims regarding compliance and security. By reducing the overhead associated with vendor management processes, organizations can achieve significant cost savings. The efficiency gained from automated assessments and real-time monitoring decreases the likelihood of costly security breaches and non-compliance penalties. The framework encourages a collaborative environment where vendors are incentivized to maintain high security and compliance standards. This can lead to stronger partnerships, as both organizations and vendors work together to ensure mutual success in navigating regulatory requirements and mitigating risks. As cyber threats evolve, the framework’s continuous monitoring and real-time data capabilities allow organizations to adapt their risk management strategies promptly. This adaptability is crucial in maintaining a robust security posture within the supply chain.

At the core of this blockchain-enhanced framework is the implementation of smart contracts, which serve as a transformative tool for automating and refining the vendor assessment process. These self-executing contracts operate based on predefined criteria, such as adherence to established security standards and the successful completion of rigorous risk assessments. By utilizing smart contracts, organizations can significantly streamline the assessment process, enhancing both efficiency and effectiveness.

One of the key advantages of smart contracts is their ability to minimize human error and bias, which are often prevalent in traditional assessment methodologies. For instance, when a vendor submits the necessary documentation for evaluation, the smart contract can automatically validate this information against established benchmarks for all security controls. If any discrepancies or deficiencies are identified during this validation, the smart contract can trigger predefined actions, such as notifying relevant stakeholders or initiating further evaluation processes. This capability ensures that only compliant vendors are allowed to proceed through the assessment pipeline, thereby fortifying the overall security posture of the organization. Additionally, the framework incorporates a continuous monitoring mechanism, empowered by blockchain's real-time data capabilities. This aspect of the framework is crucial, as it allows organizations to maintain oversight of vendor activities and security postures on an ongoing basis. By continuously monitoring vendor compliance with security requirements, organizations can swiftly identify and address any emerging vulnerabilities, ensuring that vendors not only meet initial security standards but also maintain compliance over time. This proactive approach significantly reduces the risk of data breaches and operational disruptions, which can result from lax security practices among third-party vendors.

Moreover, the transparency inherent in blockchain technology further enhances the credibility of the assessment process. Every action and transaction executed by the smart contracts is recorded on an immutable ledger, creating a transparent audit trail that can be accessed by authorized parties. This level of transparency not only fosters trust among stakeholders but also simplifies regulatory compliance by providing verifiable records of vendor assessments and security measures taken. By integrating smart contracts and real-time monitoring into the vendor assessment process, this blockchain-based approach enhances security and compliance, reduces human error, and instills greater trust in vendor relationships. Ultimately, it safeguards sensitive data and mitigates risks across the supply chain, reinforcing the organization's commitment to maintaining robust security practices.

In the next section, we will present a case study that illustrates the effective implementation of this proposed approach, showcasing its practical application and the tangible benefits it can provide in a real-world scenario. Through this case study, we aim to demonstrate how organizations can leverage blockchain technology to enhance their vendor assessment processes, leading to improved security and compliance outcomes

\section{Case Study: Enhancing Security in Healthcare IoT with Blockchain-Integrated AWS Cloud Solutions}
\label{study}

In this case study, we consider that smart healthcare IoT device companies like iHealth and 100Plus consider using AWS Cloud for data storage, analytics, and visualization, ensuring robust security is crucial due to the sensitive nature of healthcare data. Before transitioning to AWS Cloud, iHealth, a leading provider of health monitoring devices, performed a thorough third-party security assessment to evaluate its current security posture and readiness against emerging threats, including zero-day attacks. This assessment focused on identifying vulnerabilities, assessing existing security controls, conducting penetration tests, and performing comprehensive vulnerability scans. The assessment uncovered several critical weaknesses, such as outdated software, weak access controls, and inadequate real-time monitoring. These vulnerabilities heightened the risks of data breaches and operational disruptions. To mitigate these risks, the company implemented several key recommendations, including a rigorous patch management process, enhanced access controls with multi-factor authentication and role-based access, advanced threat detection technologies, and a zero trust security model. Regular security training for employees was also advised to ensure a well-rounded defense.

Incorporating blockchain technology into this security framework further enhanced the security and integrity of iHealth’s operations. Blockchain-enabled smart contracts are deployed to automate the vendor assessment process, reducing human error and improving accuracy by ensuring all security controls were consistently applied and monitored. The decentralized nature of blockchain also strengthened data integrity by providing tamper-proof logs of vendor interactions and compliance records. AWS-specific security features like Identity and Access Management (IAM), AWS Shield, and CloudTrail were leveraged to enhance monitoring, auditing, and incident response. To demonstrate how blockchain enhances third-party vendor assessment and integrates various security controls, there are few steps:

\begin{itemize}
    \item Initiating Vendor Assessment: Before starting the assessment, a vendor must authenticate its identity. Blockchain can verify vendor credentials via decentralized identity management, ensuring tamper-proof validation of the vendor’s identity. A blockchain ledger securely records vendor registration and identity information, ensuring authenticity and reducing the risk of fraudulent entities participating in the ecosystem.
    \item Submission of Vendor Documentation: Vendors submit documentation for compliance, which may include security policies, certifications, and audit reports. Each document is hashed and stored on the blockchain to ensure data integrity. This guarantees that the submitted documents are immutable and auditable, preventing tampering.

    \item Automated Smart Contract-Based Compliance Check: Blockchain smart contracts automatically verify vendor documentation and security posture against predefined standards (e.g., NIST 800-53). Smart contracts trigger actions based on predefined criteria, such as whether the vendor has met compliance standards. For example, if a vendor meets all security controls, a certificate of compliance can be issued. If discrepancies arise, additional checks are triggered.
    \item Risk Assessment: Vulnerability scans and penetration tests are conducted to evaluate the vendor's security posture. Results from vulnerability assessments are recorded on the blockchain to provide an immutable log of identified risks, past scans, and any remediation steps taken. This ensures transparency and accountability in mitigating discovered vulnerabilities.
    \item Data Access and Control Verification: Vendors are assessed on their implementation of robust access control mechanisms. Access control policies, such as MFA and role-based access control, are verified and stored on the blockchain. This ensures these policies are enforceable and traceable across multiple vendors in real time.
    \item Continuous Monitoring: Continuous monitoring ensures that vendor systems are constantly checked for anomalies or potential breaches. Blockchain provides real-time visibility into the vendor’s operational status by storing security logs and threat detection alerts in a decentralized ledger. This decentralized architecture prevents a single point of failure and ensures comprehensive oversight.
    \item Incident Response and Remediation: When a security incident occurs, a response plan is automatically initiated. Incident response actions and their timelines are recorded in the blockchain ledger, ensuring that the vendor promptly addresses issues and complies with pre-defined response time standards. Blockchain provides auditable evidence of how an incident was handled, including the remediation steps.
    
\end{itemize}

In the next section, we present the results in terms of vulnerability reduction and incident response time improvements achieved through the use of blockchain technology.

\begin{figure}
\centering
\includegraphics[width=.5\textwidth, height=.2
\textheight]{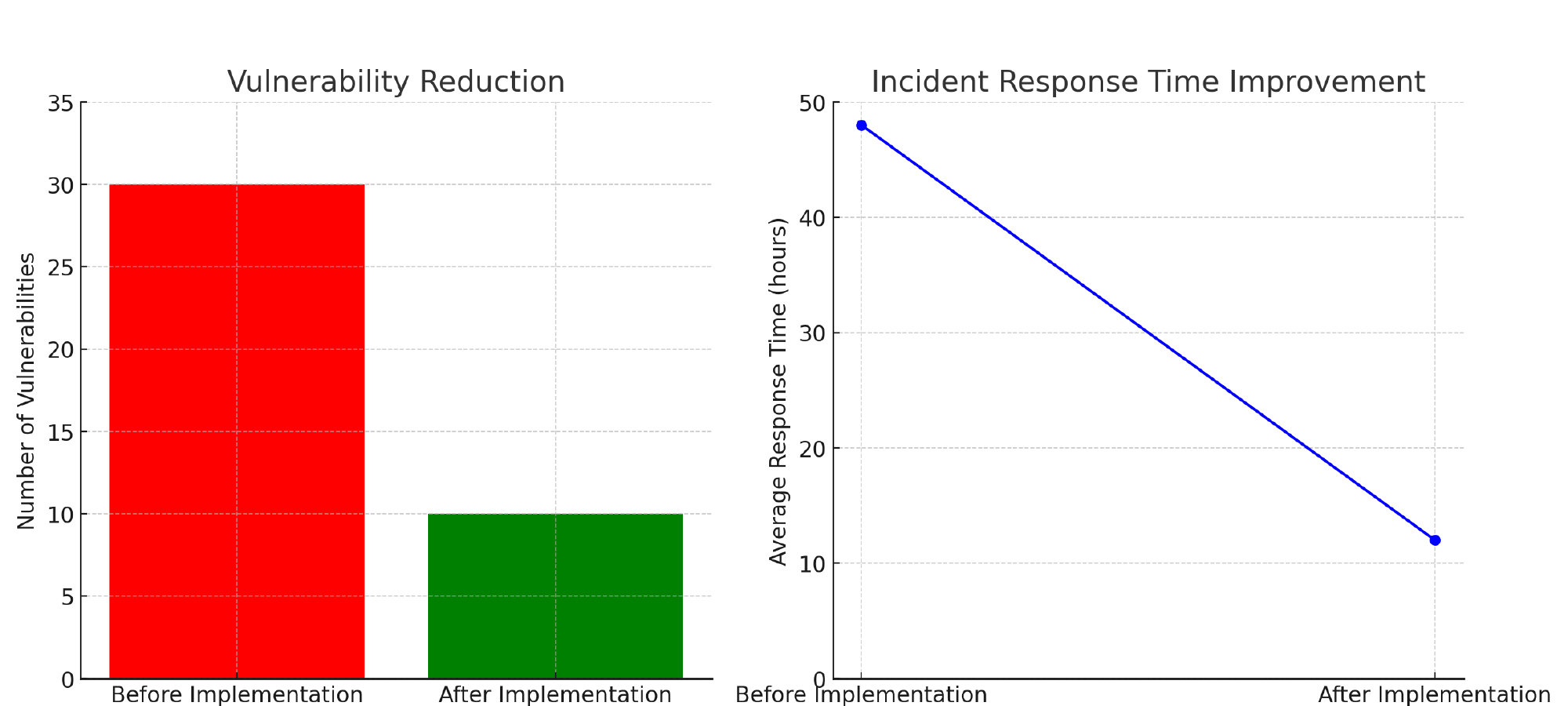}
\centering
\caption{Vulnerability Reduction and Incident Response Time Improvement by using Proposed Approach}
\label{result}
\end{figure}

\section{Results}
\label{results}
In this section, we present the results of the security assessment and the enhancements implemented for iHealth's transition to AWS Cloud. The results are categorized into two key areas: vulnerability reduction and incident response time improvement, both of which leverage blockchain technology. These findings are illustrated in Fig.~\ref{result}. We conduct this experiment using the smart health dataset~\cite{gupta2021detecting} during the transition to AWS Cloud.

\subsection{Vulnerability Reduction}

The integration of the proposed security measures, particularly those enhanced by blockchain technology, is expected to yield a significant reduction in identified vulnerabilities within the vendor assessment process. This proactive approach to security is pivotal in addressing the inherent risks associated with third-party vendors, especially in sensitive environments like healthcare. Prior to implementing these changes, a comprehensive assessment revealed a baseline of 30 vulnerabilities across various areas of security, including outdated software, weak access controls, and inadequate monitoring practices. Once the recommended security measures were applied—such as robust patch management, multi-factor authentication, role-based access control, and the adoption of a zero trust security model—this number was dramatically reduced to just 10 vulnerabilities. This 67\% reduction in vulnerabilities is a testament to the effectiveness of the new security framework and its reliance on blockchain technology. The inherent transparency and immutability of blockchain not only facilitate real-time monitoring of compliance and vulnerabilities but also ensure that any identified issues can be promptly addressed and documented. Smart contracts automate various validation processes, reducing human oversight and allowing for immediate remediation of vulnerabilities as they arise.

\subsection{Incident Response Time Improvement}

Furthermore, the implementation of advanced threat detection technologies, coupled with the principles of a zero trust security model, has resulted in a remarkable improvement in incident response times. Before the implementation of these measures, the average incident response time was 48 hours, reflecting the delays and complexities often associated with traditional security protocols. In stark contrast, the post-implementation data shows an average response time reduced to just 12 hours—a significant improvement of 75\%. This enhancement in response time is crucial for minimizing the potential impact of security incidents, particularly in a landscape where timely intervention can mean the difference between a minor issue and a catastrophic data breach.
\section{Conclusion and Future Work}

\label{conclusion} 
In conclusion, this research presents a blockchain-enhanced framework that strengthens third-party vendor risk management by integrating critical security controls. By leveraging blockchain technology, organizations can automate assessments using smart contracts, ensuring tamper-proof audits, real-time monitoring, and reducing human error in the evaluation process. This approach enhances transparency, immutability, and trust, thereby fortifying the security posture of vendors across diverse sectors. Our case study demonstrated a significant reduction in vulnerabilities and an improvement in incident response times, confirming the practical applicability of this framework. This framework also addresses key concerns such as unauthorized access, outdated software, and weak monitoring systems, ensuring vendors maintain continuous compliance with security standards.

Future research will focus on refining the blockchain-enhanced framework by incorporating advanced machine learning models to predict potential risks in real time. Additionally, efforts will be made to address scalability challenges associated with blockchain, ensuring it can handle the growing volume of vendor transactions efficiently. Expanding this research into other domains like financial services and logistics, where vendor security is equally critical, will also be a priority. Finally, conducting larger-scale case studies across multiple industries will provide deeper insights into how the framework performs under various operational conditions, further enhancing third-party risk management practice
{
\bibliographystyle{IEEEtran}
\bibliography{References}
}
\end{document}